\documentclass[pra,twocolumn]{revtex4}%
\usepackage{amsfonts}
\usepackage{amsmath}
\usepackage{amssymb}
\usepackage{graphicx}%
\providecommand{\U}[1]{\protect\rule{.1in}{.1in}}

\begin{document}
\title{A scheme for tunable quantum phase gate and effective preparation of
graph-state entanglement}
\author{Gong-Wei Lin$^{1}$}
\author{Xu-Bo Zou$^{2}$}
\author{Ming-Yong Ye$^{1}$}
\author{Xiu-Min Lin$^{1}$}
\email{xmlin@fjnu.edu.cn}
\author{Guang-Can Guo$^{2}$}
\affiliation{$^{1}$School of Physics and Optoelectronics Technology, Fujian Normal
University, Fuzhou 350007, People's Republic of China}
\affiliation{$^{2}$Key Laboratory of Quantum Information, Department of Physics, University
of Science and Technology of China, Hefei 230026}

\begin{abstract}
A scheme is presented for realizing a quantum phase gate with three-level
atoms, solid-state qubits---often called artificial atoms, or ions that share
a quantum data bus such as a single mode field in cavity QED system or a
collective vibrational state of trapped ions. In this scheme, the conditional
phase shift is tunable and controllable via the total effective interaction
time. Furthermore, we show that the method can be used for effective
preparation of graph-state entanglement, which are important resources for
quantum computation, quantum error correction, studies of multiparticle
entanglement, fundamental tests of non-locality and decoherence.

\end{abstract}

\maketitle

\section{Introduction}

Since quantum computer and quantum communication can provide more powerful
ability than the classical ones \cite{Shor,Bennett}, in the past few years, a
variety of systems are explored for the realization of practical quantum
information processors. Among them cavity quantum electrodynamics (QED) system
\cite{Sleator,Zheng} and ion trap system \cite{Cirac,Anders} are two of the
ideal candidates for quantum computation and quantum communication. The
distinctive feature of these two systems are (i) they allow the implementation
of quantum gates between any set of (not necessarily neighboring) atoms,
solid-state qubits---often called artificial atoms, or ions which share a
single mode field in cavity QED systems or a collective vibrational state of
trapped ions, (ii) the atoms or artificial atoms trapped in a high-Q cavity or
the ions trapped in a potential trap have long decoherence times \cite{Nogues,
McKeever, Maunz, Sackett,Leibfried}. Because of these advantages many efforts
have been devoted to the implementation of quantum logical gates and the
generation of entangled states \cite{Solano}. In particular, for cavity QED
system several experiments have been reported about the generation of the
Einstein-Podolsky-Rosen state \cite{Hagley} of two atoms,
Greenberger-Horne-Zeilinger (GHZ) state \cite{Rausch} of three parties (two
atoms plus one cavity mode), Schr\"{o}dinger cat state \cite{Brune}, and Fock
state \cite{Brattk} of a single-mode cavity field. Remarkably, in ion trap
system maximally up to eight ions have been entangled \cite{H H}.

In recent years, a special type of multipartite entangled states, so-called
graph states \cite{Hein,Hein1}, have become the centre of attention. They can
correspond to mathematical graphs, where each vertex represents a qubit
prepared in the state $(\left\vert 0\right\rangle +\left\vert 1\right\rangle
)/\sqrt{2}$ and each edge represents a two-qubit controlled-Z gate having been
applied to the two connected qubits. An interesting feature is that many
entanglement properties of graph states are closely related to their
underlying graphs \cite{Hein,Hein1}. Besides their thought-provoking
theoretical structure, graph states have also provided new insights into
studies of nonlocality \cite{Scarani,Otfried} and become an interesting
resource for multiparty quantum communication \cite{M. Hillery}. Special
instances of graph states are codewords of various quantum error correcting
codes \cite{Schlingema}, which are of central importance for protecting
quantum states against decoherence in quantum computation \cite{Gottesman, W.
D}. In particular, special instances of graph states have served as essential
resources for quantum computation \cite{Briegel,Varnava}. Recently, much
progress with preparation of arbitrary graph states has been made in the
linear optics regime and the optical lattice \cite{Bodiya,Kay,Clark}. In Ref.
\cite{Y. Lu}, experimental entanglement of six photons in graph states was reported.

In this paper, we propose an alternative scheme for quantum computation and
quantum state engineering based on cavity QED or ion trap. The main result
from this work is twofold: first, we propose a method for realizing a tunable
quantum phase gate. The accumulated conditional phase shift $\phi$ can vary
between $0$ and $2\pi$ by controlling the total effective interaction time.
Compared with previous protocols for the tunable quantum phase gates
\cite{Rauschenb,Zhen}, our scheme encodes two logical states of a qubit on the
two stable low energy states, and the conditional phase shifts are obtained
without any real transitions of atomic internal states or cavity-photon or
vibration-phonon population. In contrast to the scheme in Ref. \cite{X. X.},
our method can be directly extended to construct multiple--qubit entangling
gates. Second, a more important result here is that our method can be used for
effective preparation of graph-state entanglement.

\section{The fundamental model and tunable quantum phase gate with three-level
atoms}

Next we assume (without loss of generality) that N identical atoms, each
having two low levels $\left\vert 0\right\rangle $, $\left\vert 1\right\rangle
$ and a high level $\left\vert e\right\rangle $, simultaneously interact with
a single-mode cavity and a classical field, as shown in Fig. 1.
\begin{figure}[ptbh]
\begin{center}
\includegraphics[width= 1.0in]{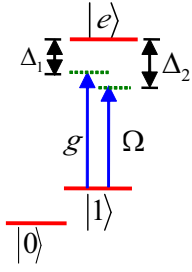}
\end{center}
\caption{(Color online) Configuration of the atomic level structure and
relevant transitions. The states $\left\vert 0\right\rangle $ and $\left\vert
1\right\rangle $ correspond to two low levels\ while $\left\vert
e\right\rangle $ is a high level. The transition $\left\vert 1\right\rangle
\rightarrow$ $\left\vert e\right\rangle $ of each atom is driven by the cavity
field and the classical pulse with the detunings $\Delta_{1}$ and $\Delta_{2}%
$, respectively. $g$ represents the coupling rate of the atom to cavity mode,
and $\Omega$ is Rabi frequency of the classical pulse.}%
\end{figure}Both of the cavity mode and the classical pulse are coupled to
$\left\vert 1\right\rangle \leftrightarrow$ $\left\vert e\right\rangle $
transition of each atom. The Hamiltonian for the whole system in the
interaction picture can be written as%

\begin{equation}
H_{int}=\sum\limits_{j=1}^{N}(g_{j}e^{i\Delta_{1}t}a\left\vert e_{j}%
\right\rangle \left\langle 1_{j}\right\vert +\Omega_{j}e^{i\Delta_{2}%
t}\left\vert e_{j}\right\rangle \left\langle 1_{j}\right\vert )+H.c.,
\end{equation}
where $a$ is the annihilation operator of the cavity mode, $\Delta_{1}$ and
$\Delta_{2}$ denote the detunings of the cavity mode and classical pulse from
respective atomic transitions, $g_{j}$ represents the coupling rate of the jth
atom to cavity mode, and $\Omega_{j}$ is Rabi frequency of the classical pulse
for the jth atom. In the case of $\Delta_{1}\gg\left\vert g_{j}\right\vert $
and $\Delta_{2}\gg\left\vert \Omega_{j}\right\vert $, the high level
$\left\vert e\right\rangle _{j}$ can be adiabatically eliminated, leading to%

\begin{align}
H_{eff}  &  =\sum\limits_{j=1}^{N}[-\frac{\left\vert g_{j}\right\vert ^{2}%
}{\Delta_{1}}a^{\dag}a\left\vert 1_{j}\right\rangle \left\langle
1_{j}\right\vert -\frac{\left\vert \Omega_{j}\right\vert ^{2}}{\Delta_{2}%
}\left\vert 1_{j}\right\rangle \left\langle 1_{j}\right\vert \nonumber\\
&  -(\lambda_{j}ae^{-i\delta t}+\lambda_{j}^{\ast}a^{\dag}e^{i\delta
t})\left\vert 1_{j}\right\rangle \left\langle 1_{j}\right\vert ],
\end{align}
where $\delta=\Delta_{2}-\Delta_{1}$, $\lambda_{j}=\Omega_{j}^{\ast}%
g_{j}(1/\Delta_{1}+1/\Delta_{2})/2$. The first and second terms in Eq. (2) are
the Stark shifts for the level $\left\vert 1_{j}\right\rangle $ that are
induced by the cavity mode and the classical pulse, respectively. The last
term is the coupling between the cavity mode and the classical field assisted
by the atoms. In the case of $\delta\gg\frac{\left\vert \Omega_{j}\right\vert
^{2}}{\Delta_{2}}$, $\frac{\left\vert g_{j}\right\vert ^{2}}{\Delta_{1}}$,
$\left\vert \lambda_{j}\right\vert $, the cavity photon is only virtually
excited and any two atoms interfere with each other \cite{Klaus}. For the sake
of convenience, we will assume $\Omega_{j}=\Omega$, $g_{j}=g$, and $\Omega$
and $g$ are real so that $\lambda_{j}=\Omega g(1/\Delta_{1}+1/\Delta
_{2})/2=\lambda$. The effective Hamiltonian is then given by%

\begin{align}
H_{eff}  &  =\sum\limits_{j=1}^{N}(-\frac{g^{2}}{\Delta_{1}}a^{\dag}%
a-\frac{\Omega^{2}}{\Delta_{2}}+\frac{\lambda^{2}}{\delta})\left\vert
1_{j}\right\rangle \left\langle 1_{j}\right\vert \nonumber\\
&  +\lambda^{^{\prime}}\sum\limits_{j,k=1,j\neq k,j<k}^{N}\left\vert
1_{j}\right\rangle \left\langle 1_{j}\right\vert \left\vert 1_{k}\right\rangle
\left\langle 1_{k}\right\vert ,
\end{align}
where $\lambda^{^{\prime}}=\frac{2\lambda^{2}}{\delta}$. The last term in Eq.
(3) describes the\ coupling between atoms j and k mediated by the cavity mode
and the classical pulse. If the cavity field is initially in the vacuum state,
the effective Hamiltonian reduces to%

\begin{align}
H_{eff}  &  =\sum\limits_{j=1}^{N}(-\frac{\Omega^{2}}{\Delta_{2}}%
+\frac{\lambda_{j}^{2}}{\delta})\left\vert 1_{j}\right\rangle \left\langle
1_{j}\right\vert \nonumber\\
&  +\lambda^{^{\prime}}\sum\limits_{j,k=1,j\neq k,j<k}^{N}\left\vert
1_{j}\right\rangle \left\langle 1_{j}\right\vert \left\vert 1_{k}\right\rangle
\left\langle 1_{k}\right\vert .
\end{align}

Suppose that the two low levels $\left\vert 0\right\rangle $ and $\left\vert
1\right\rangle $ of each atom represent two logical states of a qubit. The
time evolutions of four logical states for two qubits, under the Hamiltonian
(4), are given by%

\begin{align}
\left\vert 0_{j}\right\rangle \left\vert 0_{k}\right\rangle  &  \rightarrow
\left\vert 0_{j}\right\rangle \left\vert 0_{k}\right\rangle ,\nonumber\\
\left\vert 0_{j}\right\rangle \left\vert 1_{k}\right\rangle  &  \rightarrow
e^{-i\xi_{I}}\left\vert 0_{j}\right\rangle \left\vert 1_{k}\right\rangle
,\nonumber\\
\left\vert 1_{j}\right\rangle \left\vert 0_{k}\right\rangle  &  \rightarrow
e^{-i\xi_{I}}\left\vert 1_{j}\right\rangle \left\vert 0_{k}\right\rangle
,\nonumber\\
\left\vert 1_{j}\right\rangle \left\vert 1_{k}\right\rangle  &  \rightarrow
e^{-i(2\xi_{I}+\phi)}\left\vert 1_{j}\right\rangle \left\vert 1_{k}%
\right\rangle ,
\end{align}
where $\xi_{I}=(-\frac{\Omega^{2}}{\Delta_{2}}+\frac{\lambda^{2}}{\delta})t$
and $\phi=\lambda^{^{\prime}}t$. After the performance of the one-qubit
operation $\left\vert 1_{j(k)}\right\rangle \rightarrow$ $e^{i\xi_{I}%
}\left\vert 1_{j(k)}\right\rangle $, there is%

\begin{align}
\left\vert 0_{j}\right\rangle \left\vert 0_{k}\right\rangle  &  \rightarrow
\left\vert 0_{j}\right\rangle \left\vert 0_{k}\right\rangle ,\nonumber\\
\left\vert 0_{j}\right\rangle \left\vert 1_{k}\right\rangle  &  \rightarrow
\left\vert 0_{j}\right\rangle \left\vert 1_{k}\right\rangle ,\nonumber\\
\left\vert 1_{j}\right\rangle \left\vert 0_{k}\right\rangle  &  \rightarrow
\left\vert 1_{j}\right\rangle \left\vert 0_{k}\right\rangle ,\nonumber\\
\left\vert 1_{j}\right\rangle \left\vert 1_{k}\right\rangle  &  \rightarrow
e^{-i\phi}\left\vert 1_{j}\right\rangle \left\vert 1_{k}\right\rangle .
\end{align}
Thus a conditional phase shift $\phi=\lambda^{^{\prime}}t$ is produced if and
only if two atoms both are in the state $\left\vert 1\right\rangle $. The n
times similar operations can accumulate a phase $\phi^{^{\prime}}=%
{\displaystyle\sum\limits_{i}^{n}}
{\displaystyle\int\limits_{t_{i}}^{t_{(i+1)}}}
\lambda_{i}^{^{\prime}}dt$ varying between $0$ and $2\pi$, which is
controllable via the total effective interaction time $t$ ($t_{i}$ denotes the
beginning time of the ith operation). Ref. \cite{Bremner} shows that any
conditional quantum phase gate is universal, since all quantum computations
can be realized by combining it and rotations of individual qubits. For
example, with the choice of $t=\frac{\pi}{\lambda^{^{\prime}}}$, a two-qubit
controlled-Z gate is obtained, which is a familiar two-qubit universal logic
gate \cite{Barenco}. We note that Yi et al. have proposed a novel scheme for
conditional quantum phase gate between two 3-state atoms via simultaneously
driven by a single-mode cavity field and a very weak classical field \cite{X.
X.}. However, there are the following differences between our protocol and
that in Ref. \cite{X. X.}. (a) Our scheme does not need the wave front delay
of the classical field between the two atoms corresponds to an odd number of
$\pi$, which require that two atoms are separated from each other by at least
half optical wavelength of the classical field. (b) Our method can be extended
to construct multiple--qubit entangling gates, which are used as basic tools
for effective generation of graph-state entanglement in next section. But the
scheme in Ref. \cite{X. X.} is difficult to simultaneously operate a
controlled phase gate between two arbitrary atoms coupled to the cavity, due
to one can't make the wave front delay of the classical field between
arbitrary two of several atoms correspond to an odd number of $\pi$. (c)
Unlike the scheme in Ref. \cite{X. X.}, our scheme doesn't require weak
classical field ($\Omega<g)$ for the large detuning.

\section{Effective generation of graph-state entanglement}

We first give a brief review of the definition and properties of graph states
\cite{Hein,Hein1}. An n-qubit graph state is defined as the coeigenstate of n
independent stabilizer operators $S_{i}=X_{i}\Pi_{j}Z_{j}$, where i denotes
qubit i (each qubit is associated with a vertex of the graph), j runs over all
the neighbors of the qubit $i$, and $X_{i}$, $Z_{i}$ are simply the Pauli
operators $\sigma_{x}$ and $\sigma_{z}$ for qubit $i$. In a graph state, the
qubits $i$ and $j$ are called neighbor if they are connected with an edge. The
graph state reduces to a cluster state if the corresponding graph is a
periodic lattice \cite{Briegel}. Each graph can be represented by a diagram in
a plane, where each vertex is represented by a point and each edge by an arc
joining two not necessarily distinct vertices. In this pictorial
representation many concepts related to graphs can be visualized in a
transparent manner. A graph $\left\vert G\right\rangle $ is local unitary (LU)
equivalent representation of another graph $\left\vert G^{^{\prime}%
}\right\rangle $, if the graph $\left\vert G\right\rangle $ can be transformed
into $\left\vert G^{^{\prime}}\right\rangle $ only by local unitary operation.
It has been proved that, by local complementation of a graph $\left\vert
G\right\rangle $ at one vertex and leaving the rest of the graph $\left\vert
G\right\rangle $ unchanged, a new graph $\left\vert G^{^{\prime}}\right\rangle
$ obtained is LU-equivalent representation of the graph $\left\vert
G\right\rangle $. This is called LU rule \cite{Hein1} (or LC rule \cite{Hein}).

The graph state $\left\vert G\right\rangle $ can be obtained by applying a
sequence of commuting unitary two-qubit controlled-Z gate $U_{z}^{(a,b)}$ to
the empty graph $\left\vert +\right\rangle ^{\otimes n}$: $\left\vert
G\right\rangle $ = $\Pi_{(a,b)\in E}U_{z}^{(a,b)}\left\vert +\right\rangle
^{\otimes n}$, where $E$ denotes the set of edges in the graph $\left\vert
G\right\rangle $, and $\left\vert +\right\rangle =(\left\vert 0\right\rangle
+\left\vert 1\right\rangle )/\sqrt{2}$. The unitary two-qubit operation
$U_{z}^{(a,b)}$ on the vertices $a$, $b$ adds or removes the edge $\{a,b\}$.
In the following, with a combination of the LU rule and the multiple--qubit
entangling gates, we show an efficient scheme for preparation of graph-state
entanglement. \begin{figure}[ptbh]
\begin{center}
\includegraphics[width= 3.0in]{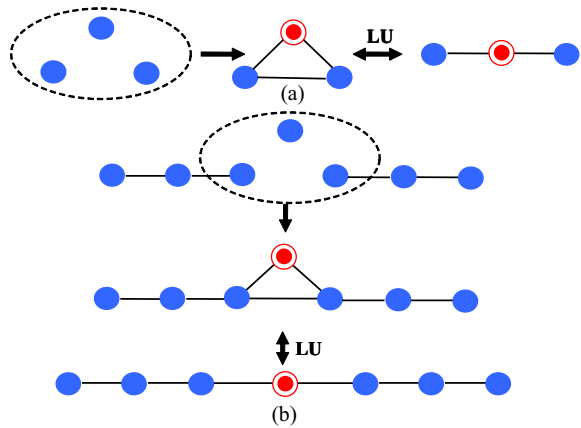}
\end{center}
\caption{(Color online) (a) The representation of the three-qubit entangling
gate. (b) Illustration of fusing two three-qubit linear cluster states into a
larger one by a three-qubit entangling gate. The ringed point denotes the
vertex, to which the LU rule is applied. The notation LU denotes that two
graph states can be transformed into each other only by local unitary
operation.}%
\end{figure}

\textit{Three-qubit entangling gate}.--- As shown in Fig. 2(a), three atoms 1,
2, and 3 with the initial state $\left\vert \Phi_{3}\right\rangle =\frac
{1}{\sqrt{2^{3}}}\Pi_{j=1}^{3}(\left\vert 0\right\rangle _{j}+\left\vert
1\right\rangle _{j})$ trapped in a cavity are simultaneously driven by the
cavity mode and an appropriate classical pulse (Fig. 1). The state evolution
of the three-atom system is governed by%

\begin{equation}
H_{eff}=\lambda^{^{\prime}}\sum\limits_{\text{ }j\neq k,j<k}\left\vert
1_{j}\right\rangle \left\langle 1_{j}\right\vert \left\vert 1_{k}\right\rangle
\left\langle 1_{k}\right\vert ,
\end{equation}
where $j,k=1,2,3$, $\lambda^{^{\prime}}=\frac{2\lambda^{2}}{\delta}$, and the
self-energy terms have been assumed to be corrected by the added classical
pulses \cite{Rauschen}. After the effective interaction time $t=\frac{\pi
}{\lambda^{^{\prime}}}$, a three-qubit entangling gate is operated on the
three-atom system and a graph state%

\begin{equation}
\left\vert \Phi_{3}^{^{\prime}}\right\rangle =U_{z}^{(1,2)}U_{z}^{(2,3)}%
U_{z}^{(1,3)}\left\vert \Phi_{3}\right\rangle ,
\end{equation}
is obtained. The graph state is LU-equivalent representation of the
three-qubit linear cluster. Fig. 2(b) shows that two three-qubit linear
clusters are fused into a seven-qubit graph state by a three-qubit entangling
gate, which is LU-equivalent representation of a seven-qubit linear cluster.
With sequentially applying the method in Fig. 2, we can get n-qubit (n is odd)
linear cluster state through $\frac{n-1}{2}$times successful three-qubit
entangling gates.

\begin{figure}[ptbh]
\begin{center}
\includegraphics[width= 3.3in]{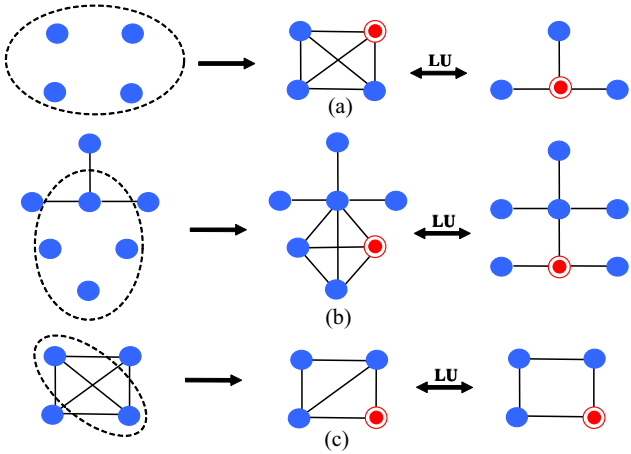}
\end{center}
\caption{(Color online) (a) The representation of the four-qubit entangling
gate. (b) Illustration of using the four-qubit entangling gates to generate a
H-sharp graph state. (c) Illustration of constructing a box graph by
performing an added controlled-Z gate on two diagonal vertexes of the graph in
(a). Other notations are the same as Fig. 2.}%
\end{figure}

\textit{Four-qubit entangling gate}.--- In similar way, as sketched in Fig.
3(a), four atoms 1, 2, 3, and 4 with the initial state $\left\vert \Phi
_{4}\right\rangle =\frac{1}{\sqrt{2^{4}}}\Pi_{j=1}^{4}(\left\vert
0\right\rangle _{j}+\left\vert 1\right\rangle _{j})$ simultaneously interact
with a cavity mode and an appropriate classical pulse (Fig. 1). After the
effective interaction time $t=\frac{\pi}{\lambda^{^{\prime}}}$, the state
evolution of four-atom system is%

\begin{equation}
\left\vert \Phi_{4}^{^{\prime}}\right\rangle =U_{z}^{(1,2)}U_{z}^{(1,3)}%
U_{z}^{(1,4)}U_{z}^{(2,3)}U_{z}^{(2,4)}U_{z}^{(3,4)}\left\vert \Phi
_{4}\right\rangle ,
\end{equation}
which is a four-qubit graph state. The graph state is LU-equivalent
representation of a four-qubit star graph, which is equivalent to
four-particle GHZ state \cite{Hein,Hein1}. Fig. 3(b) shows that with two
four-qubit entangling gates one can obtain a seven-qubit graph state, which is
LU-equivalent representation of an H-sharp graph. Fig. 3(c) shows that after
performing an controlled-Z gate on two diagonal vertexes of the graph in Fig.
3(a) \cite{111}, we can obtain another class graph state, which is
LU-equivalent representation of a box graph.

\begin{figure}[ptbh]
\begin{center}
\includegraphics[width= 6.6in]{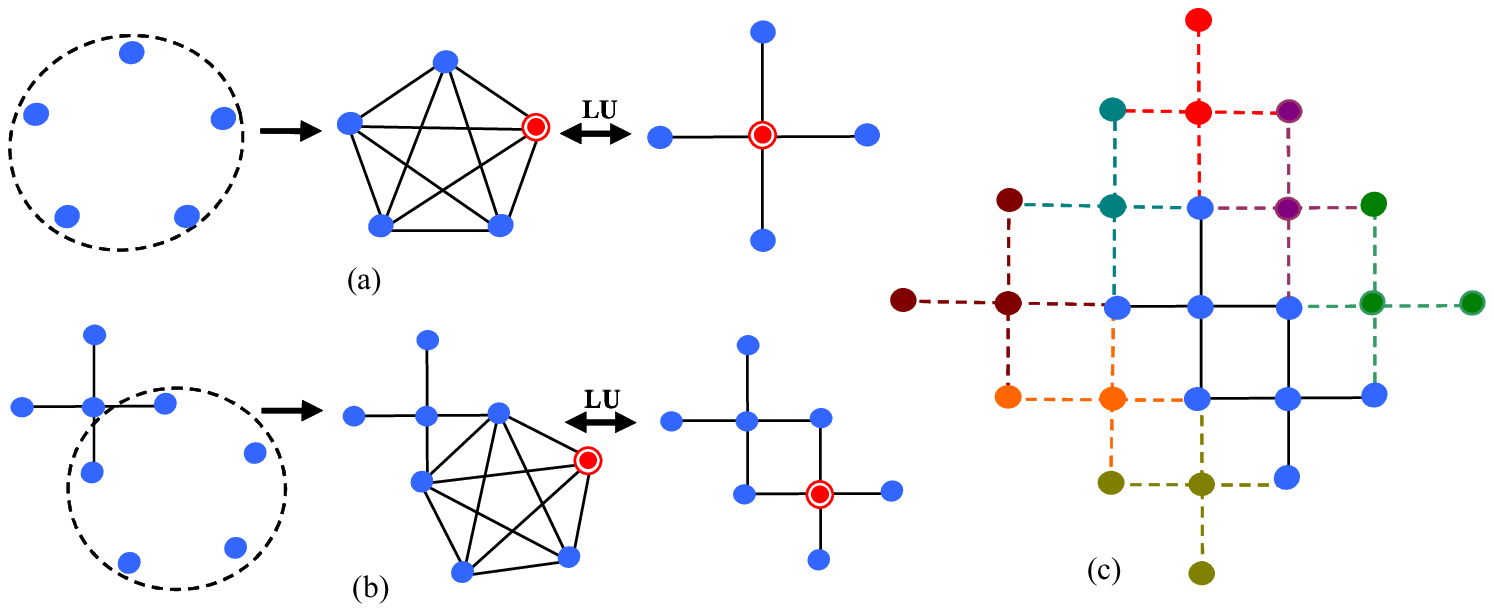}
\end{center}
\caption{(Color online) (a) The representation of the five--qubit entangling
gate. (b) and (c) Illustration of using the five-qubit entangling gates to
generate square lattice cluster states. Other notations are the same as Fig.
2.}%
\end{figure}

\begin{figure}[ptbh]
\begin{center}
\includegraphics[width= 6.8in]{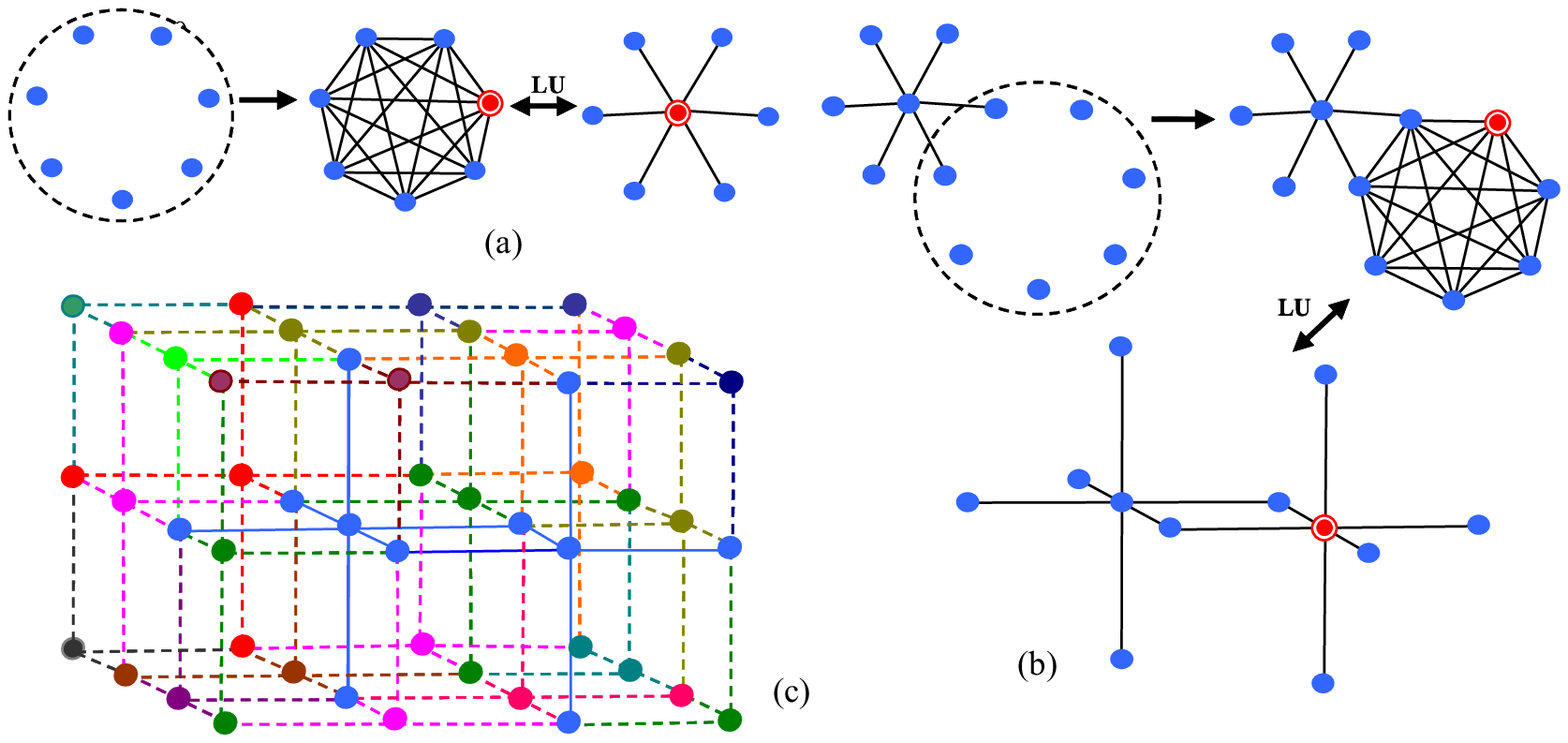}
\end{center}
\caption{(Color online) (a) The representation of the seven--qubit entangling
gate. (b) and (c) Illustration of using the seven-qubit entangling gates to
generate 3D cluster states. Other notations are the same as Fig. 2.}%
\end{figure}

\textit{Five or more-qubit entangling gate}.---Five or more atoms
simultaneously interact with a cavity mode and an appropriate classical pulse.
After the effective interaction time $t=\frac{\pi}{\lambda^{^{\prime}}}$, we
obtain a five or more-qubit entangling gate. As sketched in Fig. 4(a), after
applying a five-qubit entangling gate, one can get a five-qubit graph sate,
which is LU-equivalent representation of a five-qubit star graph state being
equivalent to a five-particle GHZ state \cite{Hein,Hein1}. Fig. 4(b) shows
that a eight-qubit 2D graph state can be obtianed by applying two five-qubit
entangling gates. Fig. 4(c) shows how to effectively generate 2D square
lattice cluster states by using the five-qubit entangling gates, which can be
used as the essential resource for realization of universal quantum
computation \cite{Briegel}. In Fig. 5, we illustrate a method for effective
preparation of three-dimensional (3D) cluster state by sequentially using the
seven-qubit entangling gates. In Ref. \cite{11}, Raussendorf et al. have shown
that 3D cluster state is a fault-tolerant fabric of one-way quantum computer.

Here we only display a sample of generating some typical graph states. We
believe that it is still effective to generate a more complex graph state by
combining the LU rule, the multiple--qubit entangling gates, and any two-qubit
controlled-Z gate.

\section{Discussion and conclusion}

The required atomic level configuration and transitions can be achieved in
neutral atoms trapped in an optical cavity, solid-state qubits---often called
artificial atoms trapped in a microwave cavity, or the ions trapped in the
potential trap. It should be noted that one needs to reach the strong-coupling
limit and the Lamb-Dicke regime in order to perform the gate successfully. In
optical cavity QED system, strong-coupling limit and localization to the
ground state of motion for an atom trapped in an optical cavity are still a
challenging pursuit \cite{Wilk}. Conversely, for the solid-state system or
ion-trap system, strong-coupling and Lamb-Dicke limit can be easier to achieve
\cite{Wallraff}. In the past few years, much attention has been paid to
superconducting devices such as Cooper pair boxes, Josephson junctions, and
superconducting quantum interference devices (SQUIDs), since these solid-state
artificial atoms are relatively easy to scale up and have been demonstrated to
have relatively long decoherence time \cite{Vion,Y. Yu}. Recently, for SQUID
system, many methods for realizing a single-qubit arbitrary rotation gate and
a two-qubit controlled-NOT (or controlled-Z gate) have been presented \cite{P.
Yang,Yang}. Any particular SQUID qubit operation can be realized by adjusting
level structure of individual qubit \cite{P. Yang}. In the following we first
discuss that it is possible to experimentally demonstrate our scheme with the
SQUIDs in a microwave cavity.

Typically, for SQUIDs interacting with a high-Q microwave cavity , the
coupling constant is $g\sim1.8\times10^{8}Hz$, the photon lifetime (the energy
relaxation time of excited state $\left\vert e\right\rangle $) is
$t_{c(r)}\sim7.6\times10^{-7}s$ \cite{P. Yang}. With the choices of
$\Omega=1.05g$, $\Delta_{1}=20g$, $\Delta_{2}=21g$, we have $\delta=\Delta
_{2}-\Delta_{1}=g$\ $\sim20\frac{g^{2}}{\Delta_{1}}$, $19.05\frac{\Omega^{2}%
}{\Delta_{2}}$, $19.5\lambda$, which satisfy the conditions $\Delta_{1}\gg g$,
$\Delta_{2}\gg\Omega$, and $\delta\gg\frac{g^{2}}{\Delta_{1}}$, $\frac
{\Omega^{2}}{\Delta_{2}}$, $\lambda$. Our calculations show that (i) the
occupation probability of the excited state $\left\vert e\right\rangle $ is
$P_{r}$ $\sim0.01$ ($\simeq\frac{\Omega^{2}}{\delta^{2}}$), thus the effective
energy relaxation time is $t_{r}^{^{\prime}}\sim P_{r}^{-1}t_{r}\sim76\mu s$;
(ii) the occupation probability of the photon is $P_{c}$ $\sim0.01$
($\simeq\frac{\lambda^{2}}{\delta^{2}}$), thus the effective photon lifetime
is $t_{c}^{^{\prime}}\sim P_{c}^{-1}t_{c}\sim76\mu s$; (iii) the required
effective interaction time for a two-qubit controlled-Z gate or a
multiple--qubit entangling gate is $t\sim\frac{\pi}{\lambda^{^{\prime}}}%
\sim3\mu s$. Therefore, it is possible to perform\ several two-qubit
controlled-Z gates or multiple--qubit entangling gates within the decoherence
time $t_{r}^{^{\prime}}\sim t_{c}^{^{\prime}}\sim76\mu s$.

In the ion-trap system, several ions are assumed to be confined in a linear
trap \cite{Cirac,Anders} and have been cooled to the ground state. The ions
are simultaneously excited by two laser fields (for operating a controlled
phase gate between two or several arbitrary ions, one needs single laser
pulses to address two or several arbitrary ions individually, which have been
demonstrated in the experiment \cite{Schmidt}). One is (off-resonantly) tuned
to the first lower vibrational side and the other is insensitive to the motion
with appropriate frequency \cite{X. X.}, as shown in Fig. 1. In the Lamb-Dicke
limit the Hamiltonian of the system is then again described by Eq. (1), except
now $a$ is the annihilation operator of the collective vibrational mode
instead of the cavity mode. Considering experimental parameters of ion
experiments at Ref. \cite{Sackett}, we can achieve an effective coupling rate
$\lambda^{^{\prime}}\sim10^{4}Hz$. So the required effective interaction time
for a two-qubit controlled-Z gate or a multiple--qubit entangling gate is
$t\sim0.1ms$, shorter than the typical motional decoherence time $t_{d}$
$\sim10ms.$

In summary, we have proposed a scheme to realize a tunable quantum phase gate
with three-level atoms or artificial atoms or ions simultaneously interacting
with a quantum data bus such as a single mode field in cavity QED system, or a
collective vibrational state of trapped ions. The distinct advantages of our
scheme are that there is no cavity-photon population involved and the atoms
are always in their low levels. In addition, our method can be extended to
construct multiple--qubit graph-state entanglement. The method opens up a
prospect to generate large-scale graph state in cavity QED system and ion trap system.

\textbf{Acknowledgments:} This work was funded by National Natural Science
Foundation of China (Grant No. 10574022), the Natural Science Foundation of
Fujian Province of China (Grant No. 2007J0002), and the Foundation for
Universities in Fujian Province (Grant No. 2007F5041).

\bigskip


\begin{thebibliography}{99}
\bibitem {111}For operating a controlled phase gate between two arbitrary
atoms trapped in the cavity, one need the atoms are separated by at least one
optical wavelength so that single laser pulses can address each atom
individually, which is difficult to achieve with the current experimental technology.
\end{thebibliography}
\end{document}